\documentstyle[prd,aps,epsf,preprint]{revtex}

\begin{document}

\preprint{                                                 BARI-TH/324-98}
\draft
\title{  Super-Kamiokande data and atmospheric neutrino decay	}
\author{         G.~L.~Fogli, E.~Lisi, A.~Marrone, and G.~Scioscia}
\address{   Dipartimento di Fisica and Sezione INFN di Bari, \\
                  Via Amendola 173, I-70126 Bari, Italy}
\maketitle
\begin{abstract}
%...........................................................................
Neutrino decay has been proposed as a possible solution to the atmospheric
neutrino anomaly, in the light of the recent data from the Super-Kamiokande
experiment. We investigate this  hypothesis by means of a quantitative analysis
of the zenith angle  distributions of neutrino events in Super-Kamiokande,
including the latest (45 kTy) data.  We find that the neutrino decay hypothesis
fails to reproduce the observed distributions of muons. 
%...........................................................................
\end{abstract}

The Super-Kamiokande (SK) experiment has confirmed, with high statistical
significance, the anomalous flavor composition of the  atmospheric neutrino
flux. Such anomaly  is found in all SK data samples, including (in order of
increasing energy) sub-GeV  $e$-like and $\mu$-like events (SG$e$ and SG$\mu$)
\cite{SKSG}, multi-GeV  $e$-like  and $\mu$-like events (MG$e$ and MG$\mu$)
\cite{SKMG}, and upward through-going muons (UP$\mu$) \cite{SKUP}. In
particular, all muon event samples (SG$\mu$, MG$\mu$, and UP$\mu$) show
significant distortions of the observed zenith angle distributions, as compared
with standard expectations. The recent muon data from the MACRO \cite{MACR} and
Soudan-2  \cite{SOUD} experiments, as well as from the finalized Kamiokande
sample \cite{KAUP}, are also consistent with the Super-Kamiokande data.

The observed dependence of the muon deficit on both energy and direction can be
beautifully explained  via neutrino flavor oscillations in the
$\nu_\mu\to\nu_\tau$ channel \cite{EVID}. Transitions into sterile states are
also consistent with the data \cite{Mark,Alec},  as well as subdominant
oscillations in the $\nu_\mu\to\nu_e$ channel \cite{LISI}.  Determing the
flavor(s) of the oscillating partner(s) of the muon neutrino  will represent a
crucial test of such explanation(s). In the meantime, it is useful to challenge
the oscillation hypothesis and to investigate possible alternative scenarios
\cite{LOSE}.

Neutrino decay  has been recently proposed  \cite{DECA} as a possible solution
to the atmospheric neutrino anomaly.  In a nutshell, the muon neutrino
$\nu_\mu$  is assumed to have an unstable (decaying) component $\nu_d$,
%.........................................................................
\begin{equation} 
\cos\xi \equiv \langle \nu_\mu | \nu_d \rangle \neq 0\ ,
\end{equation} 
%...........................................................................
with mass and lifetime $m_d$ and $\tau_d$, respectively.  In the parameter
range of interest, the $\nu_\mu$ survival probability reads \cite{DECA} 
%...........................................................................
\begin{equation} 
P_{\mu\mu} \simeq \sin^4\xi + \cos^4\xi
\exp(-\alpha L/E)\ ,  
\label{pmumu} 
\end{equation} 
%...........................................................................
where $\alpha=m_d/\tau_d$, and $L/E$ is the ratio between the neutrino
pathlength and energy. The possible unstable component of $\nu_e$ is 
experimentally constrained to be very small \cite{DECA,PDGr}, so one can take
$P_{ee}\simeq 1$.

For large values of $\cos\xi$ and for $\alpha\sim O(D_\oplus/1 {\rm\ GeV})$
($D_\oplus=12,800$ km), the exponential term in Eq.~(\ref{pmumu}) can produce a
detectable  modulation of the muon-like event distributions \cite{DECA}. In
particular, the expected modulation seems to be roughly in agreement
\cite{DECA} with the reconstructed $L/E$ distribution of  contained SK events
\cite{EVID}. However, the $L/E$ distribution is not really suited to
quantitative tests, since it is affected by relatively large uncertainties,
implicit in backtracing the (unobservable) parent neutrino momentum vector from
the  (observed) final lepton momentum. Moreover, the $L/E$ distribution
includes only a fraction of the data (the fully contained events), and  mixes
low-energy and high-energy events, making it difficult to judge how the
separate data samples (SG, MG, and UP) are fitted by the decay solution. 
Therefore, we think it worthwhile to  test the neutrino decay hypothesis in a
more convincing and  quantitative way, by using observable quantities (the
zenith distributions of the observed leptons) rather than unobservable,
indirect parameters such as $L/E$.

To this purpose we use,  as described in \cite{LISI},  five zenith-angle
distributions of neutrino-induced lepton events in Super-Kamiokande, namely,
SG$e$ and SG$\mu$ (5+5 bins),  MG$e$ and MG$\mu$ (5+5 bins),  and UP$\mu$ (10
bins), for a total of 30 data points.  The data refer to the preliminary 45
kiloton-year sample of Super-Kamiokande, as taken from \cite{Mark,Alec}. The
theoretical calculations are performed with the same technique as in
\cite{LISI}, but the flavor survival probability  refer now to $\nu$ decay
[Eq.~(\ref{pmumu})] rather than $\nu$ oscillations.  As in \cite{LISI},
conservative errors are assumed not only for the overall normalization of the 
expected distributions, but also for their shape distortions.  The
(dis)agreement between data and theory is quantified through a $\chi^2$
statistic, which takes into account the strong correlations between
systematics.

Figure~1 shows the results of our $\chi^2$ analysis in the plane  $(\cos\xi,
\alpha)$, with $\alpha$ given in unit of GeV$/D_\oplus$. The regions at
90 and 99\% C.L.\ are defined by $\chi^2-\chi^2_{\min}=4.61$ and 9.21,
respectively. As qualitatively expected, the data prefer $\alpha\simeq 1$ and
large  $\cos\xi$, in order to produce a large suppression of $\nu_\mu$'s coming
from below. However, the absolute $\chi^2$ is always much higher than the
number of degrees of freedom, $N_{DF}=28$ (30 data points $-$ 2 free
parameters). In fact,  it is  $\chi^2_{\min}=86.2$ at the best-fit point
[reached for $(\cos\xi,\alpha)=(0.95,0.90)$]. The very poor global fit
indicates that the zenith distributions cannot be accounted for by neutrino
decay, contrarily to the claim of \cite{DECA}, which was based on reduced and
indirect data (the $L/E$ distribution).  This situation should be contrasted
with the $\nu_\mu\leftrightarrow\nu_\tau$ oscillation hypothesis, which gives
$\chi^2_{\min}/N_{\rm DF}\sim 1$ both in two-flavor \cite{EVID} and
three-flavor scenarios \cite{LISI}. Even using ``older'' 
Super-Kamiokande data (i.e., the published 33 kTy sample 
\cite{SKSG,SKMG,SKUP}), we get $\chi^2_{\min}\sim 65$ for the $\nu$ decay fit,
still much higher than $N_{\rm DF}$.

Basically, neutrino decay fails to reproduce the  SK zenith distributions for
the following reason: The lower the energy, the faster the decay, the stronger
the muon deficit~---~a~pattern not supported by the data.  This can be better
appreciated in Fig.~2, which shows the SK data (45 kTy)  and the expectations 
(at the ``best-fit'' point) for the five zenith  angle $(\theta)$ 
distributions considered in our analysis. In each  bin, both the observed and
the expected lepton rates $R$ are normalized to the standard values in the
absence of decay $R_0$, so that ``no decay'' corresponds to $R/R_0=1$. The data
are shown as dots with $1\sigma$ error bars, while the decay predictions are
shown as solid lines. The predictions are affected by strongly correlated
errors (not shown), as discussed in Appendix~B of \cite{LISI}.

In Fig.~2, the muon data show significant deviations from the reference
baseline $R/R_0=1$, most notably for MG$\mu$ events. The $\nu$ decay
predictions also show some (milder) deviations, but their agreement with the
data is poor, both in normalization and in shape. Neutrino decay implies a muon
deficit decreasing with energy, at variance with the fact that the overall
($\theta$-averaged) deficit is about the same for both SG$\mu$'s and MG$\mu$'s
($-30\%$).   The shape distortions of the muon zenith distributions are also
expected to be exponentially weaker at higher energies (i.e., for slower
neutrino decay). This makes it impossible  to fit at the same time the
distorted muon distributions observed at low energy (SG$\mu$) and at high
energy (UP$\mu$), and to get a strong up-down asymmetry at intermediate energy
(MG$\mu$) as well. Notice in Fig.~2 that, although  the predicted {\em shape\/}
of the zenith distribution appears to be   in qualitative agreement with the
data pattern for SG$\mu$'s,  it is not sufficiently up-down asymmetric for
MG$\mu$'s, and it is definitely too flat for UP$\mu$'s, where the muon
suppression reaches the plateau $P_{\mu\mu}\simeq c^4_\xi+s^4_\xi$, in
disagreement with the observed $\cos\theta$-modulation \cite{Alec}.  Finally,
we remark that the electron neutrinos, despite being ``spectators'' in the
$\nu$ decay scenario (flat SG$e$ and MG$e$  distributions in Fig.~2), play a
role in the fit through the constraints on their overall rate normalization,
as in the $\nu_\mu\to\nu_\tau$ oscillation case \cite{LISI}.

In conclusion, we have shown  quantitatively  that the neutrino decay
hypothesis, although intriguing, fails to reproduce the zenith angle
distributions of the Super-Kamiokande sub-GeV, multi-GeV, and upgoing muon data
for any  value of the decay parameters $\alpha$ and $\cos\xi$. Even at the
``best fit'' point, data and expectations differ both in  total rates and in
zenith distribution shapes. Therefore, neutrino decay (at least in its simplest
form \cite{DECA}) is not a viable explanation of the Super-Kamiokande
observations. The strong disagreement between data and theory  was not apparent
in \cite{DECA}, presumably  because the experimental information used there was
rather reduced and indirect.

{\em Note added.} When this work was being completed, our attention was brought
to the recent paper \cite{LiLu}, where several scenarios---alternative to
neutrino oscillations---are considered, including neutrino decay (for
$\cos\xi=1$). The authors of \cite{LiLu}  find that neutrino decay provides a
very poor fit to the Super-Kamiokande data, consistently with our conclusions.
As far as the neutrino decay hypothesis is concerned, our work has the
advantage of being more general ($\cos\xi$ unconstrained), more refined in the
statistical approach, and updated with latest SK data (45 kTy)
\cite{Mark,Alec}.

%%%%%%%%%%%%%%%%%%%%%%%%%%%%%%%%%%%%%%%%%%%%%%%%%%%%%%%%%%%%%%%%%%%%%%%%%%%%%%%%
% 			R E F E R E N C E S 
%%%%%%%%%%%%%%%%%%%%%%%%%%%%%%%%%%%%%%%%%%%%%%%%%%%%%%%%%%%%%%%%%%%%%%%%%%%%%%%%

\begin{figure}
\vspace*{-1cm}
\epsfysize=23truecm
\epsfbox{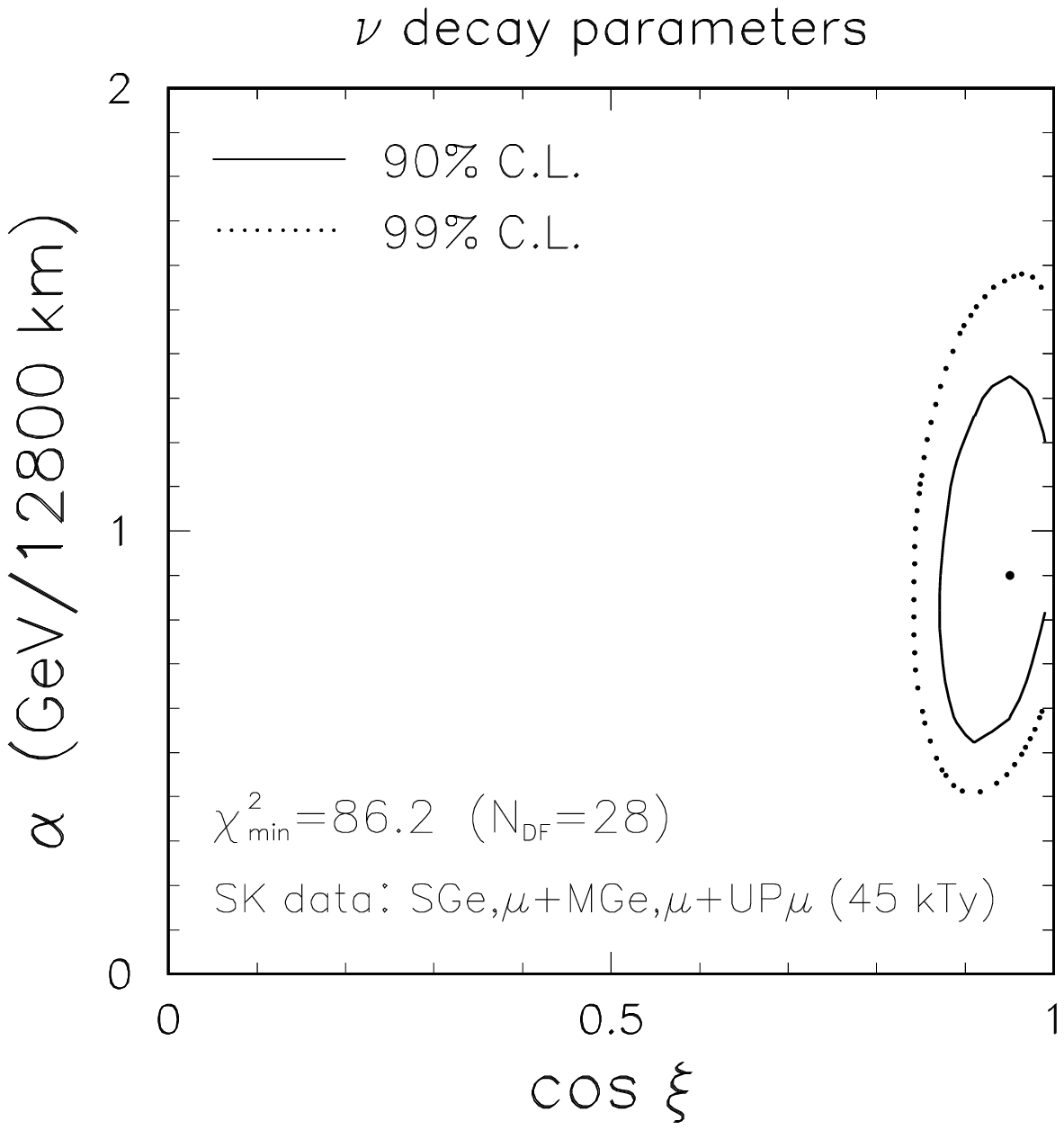}
\vskip-5cm
\caption{Fit to the Super-Kamiokande data (45 kTy, 30 data points)  in the
plane of the neutrino decay parameters  $\cos\xi=\langle \nu_\mu | \nu_d
\rangle$ and $\alpha=m_d/\tau_d$. The solid and dotted lines are defined by
$\chi^2-\chi^2_{\min}=4.61$ and 9.21, corresponding to 90\% and 99\% C.L. for
two variables. The analysis favors $\alpha\sim 1$ GeV/$D_\oplus$ and large
$\cos\xi$. However, even at the best fit point there is  poor agreement between
data and theory ($\chi^2_{\min}/N_{\rm DF}=86.2/28=3.1$), indicating that $\nu$
decay is not a viable explanation of the Super-Kamiokande observations.}
\label{fig1}
\end{figure}

%%%%%%%%%%%%%%%%%%%%%%%%%%%%%%%%%%%%%%%%%%%%%%%%%%%%%%%%%%%%%%%%%%%%%%%%%%%%%

\begin{figure}
\vspace*{-1cm}
\epsfysize=22truecm
\epsfbox{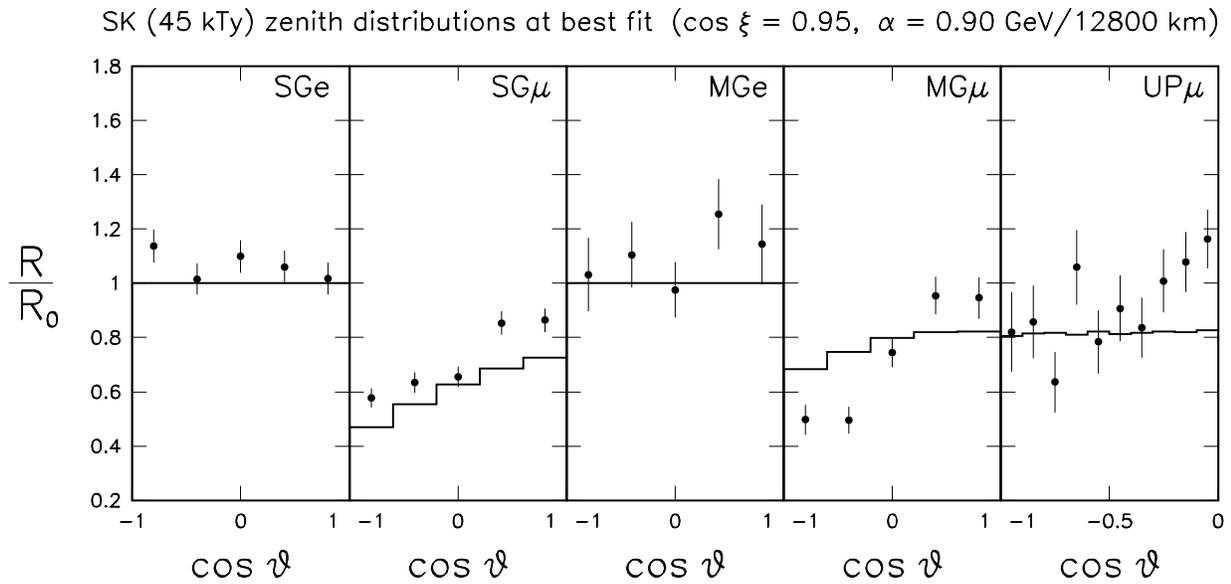}
\vskip-5cm
\caption{Zenith angle distributions of Super-Kamiokande sub-GeV $e$-like and
$\mu$-like events (SG$e$ and SG$\mu$), multi-GeV $e$-like and $\mu$-like events
(MG$e$ and MG$\mu$), and upward-going muons (UP$\mu$). Data: dots with $\pm 1
\sigma$ statistical error bars. Theory ($\nu$ decay best fit): solid curves. In
each bin, both theoretical and experimental rates $R$ are normalized to their
standard (no decay) expectations $R_0$. The solid curves do not appear to
reproduce the muon data pattern.}
\label{fig2}
\end{figure}

\end{document}